\documentclass[fleqn,usenatbib,useAMS]{mnras}

\usepackage{graphicx,graphics,placeins,float}
\usepackage{amsmath}	% Advanced maths commands
\usepackage{amssymb}	% Extra maths symbols
\usepackage{multicol}        % Multi-column entries in tables
\usepackage{bm}		% Bold maths symbols, including upright Greek
\usepackage{pdflscape}	% Landscape pages

\usepackage[authoryear]{natbib}
\usepackage{longtable,lscape,rotating}
\usepackage{supertabular,caption,multicol}
\usepackage{threeparttable}
\usepackage[T1]{fontenc}
\usepackage{ae,aecompl}

\usepackage{newtxtext,newtxmath}
\def\swift{{\em Swift}}
\def\fermi{{\em Fermi}}

\title[ATCA rapid-response observations of GRB 181123B]{Rapid-response radio observations of short GRB 181123B with the Australia Telescope Compact Array}

\author[G. E. Anderson et al.]{G. E. Anderson$^{1}$\thanks{Contact e-mail: \href{mailto:gemma.anderson@curtin.edu.au}{gemma.anderson@curtin.edu.au}}\thanks{Present address: International Centre for Radio Astronomy Research, Curtin University, GPO Box U1987, Perth, WA 6845, Australia}
M. E. Bell,$^2$
J. Stevens,$^3$
M. D. Aksulu,$^4$
J. C. A. Miller-Jones,$^1$
\newauthor
A. J. van der Horst,$^{5,6}$
R. A. M. J. Wijers,$^4$
A. Rowlinson,$^{4,7}$
A. Bahramian,$^1$
\newauthor
P. J. Hancock,$^1$
J.-P. Macquart,$^1$
S. D. Ryder,$^{8,9}$
R. M. Plotkin$^{10}$
\newauthor
\\
$^{1}$International Centre for Radio Astronomy Research, Curtin University, GPO Box U1987, Perth, WA 6845, Australia \\
$^{2}$University of Technology Sydney, 15 Broadway, Ultimo NSW 2007, Australia \\
$^3$CSIRO Astronomy and Space Science, Australia Telescope National Facility, Box 76, Epping, NSW 1710, Australia \\
$^4$Anton Pannekoek Institute for Astronomy, University of Amsterdam, Science Park 904, NL-1098 XH Amsterdam, The Netherlands \\ 
$^{5}$ Department of Physics, The George Washington University, 725 21st Street NW, Washington, DC 20052, USA\\
$^{6}$ Astronomy, Physics, and Statistics Institute of Sciences (APSIS), 725 21st Street NW, Washington, DC 20052, USA\\
$^{7}$ ASTRON, the Netherlands Institute for Radio Astronomy, Postbus 2, NL-7990 AA Dwingeloo, The Netherlands\\
$^8$ Department of Physics and Astronomy, Macquarie University, Sydney NSW 2109, Australia \\
$^9$ Macquarie University Research Centre for Astronomy, Astrophysics \& Astrophotonics, Sydney, NSW 2109, Australia\\
$^{10}$ Department of Physics, University of Nevada, Reno, NV 89557, USA
}
\date{Last updated 2021 Mar 3; in original form 2020 Apr 29}

\pubyear{2021}

\begin{document}
\label{firstpage}
\pagerange{\pageref{firstpage}--\pageref{lastpage}}
\maketitle

\begin{abstract}
We introduce the Australia Telescope Compact Array (ATCA) rapid-response mode by presenting the first successful trigger on the short-duration gamma-ray burst (GRB) 181123B. 
Early-time radio observations of short GRBs may provide vital insights into the 
radio afterglow properties of Advanced LIGO- and Virgo-detected gravitational wave events, which will in turn inform follow-up strategies to search for counterparts within their large positional uncertainties. 
The ATCA was on target within 12.6\,hr post-burst, when the source had risen above the horizon. 
While no radio afterglow was detected during the 8.3\,hr observation, we obtained 
force-fitted flux densities of $7 \pm 12$ and $15 \pm 11\mu$Jy at 5.5 and 9\,GHz, respectively. 
Afterglow modelling of GRB 181123B showed that the addition of the ATCA force-fitted radio flux densities to the \swift{} X-ray Telescope detections provided more stringent 
constraints on the fraction of thermal energy in the electrons 
(log\,$\epsilon_e = -0.75^{+0.39}_{-0.40}$ rather than log\,$\epsilon_e = -1.13^{+0.82}_{-1.2}$
derived without the inclusion of the ATCA values), which is consistent with the range of typical $\epsilon_e$ derived from GRB afterglow modelling. 
This allowed us to predict that the forward shock may have peaked in the radio band $\sim10$\,days post-burst, producing detectable radio emission $\gtrsim3-4$\,days post-burst. 
Overall, we demonstrate the potential for extremely rapid radio follow-up of transients and the importance of triggered radio observations for constraining GRB blast wave properties, regardless of whether there is a detection, via the inclusion of force-fitted radio flux densities in afterglow modelling efforts. 
\end{abstract}

\begin{keywords}

transients: gamma-ray bursts -- gamma-ray burst: individual: GRB 181123B -- radio continuum: transients -- transients: neutron star mergers -- gamma-ray burst: general

\end{keywords}

%%%%%%%%%%%%%%%%%%%%%%%%%%%%%%%%%%%%%%%%%%%%%%%%%%

%%%%%%%%%%%%%%%%% BODY OF PAPER %%%%%%%%%%%%%%%%%%

\section{Introduction}\label{sec:intro}

The first gravitational wave (GW) detection of a binary neutron star (BNS) merger, GW170817, by the Advanced
LIGO and Virgo (aLIGO/Virgo) facilities, was an eagerly awaited event \citep{abbott17a}. Such a merger was predicted to produce a multi-wavelength electromagnetic afterglow radiating from radio to gamma-rays, and GW170817 did not disappoint \citep{abbott17b,andreoni17}. The most conspicuous predicted counterpart was the prompt ejection of collimated, short-lived gamma-ray emitting jets, similar to the observed 
short gamma-ray burst \citep[SGRB, gamma-ray durations $<2$\,s;][]{narayan92} phenomenon, which is one of the two main classes of gamma-ray burst \citep[GRB;][]{norris84,dezalay92,kouveliotou93} detected by the \textit{Neil Gehrels Swift Observatory} \citep[hereafter \swift;][]{gehrels04} Burst Alert Telescope (BAT) and the \textit{Fermi Gamma-ray Space Telescope} (hereafter \fermi{}) Gamma-ray Burst Monitor \citep[GBM;][]{meegan09}. 
The other dominant population known as long GRBs (LGRBs) typically have durations $>2$\,s and are attributed to core collapse supernovae \citep[e.g.][]{galama98,bloom98}.
It was therefore the near-simultaneous detection of GW170817 and GRB 170817A \citep[the latter of which was a SGRB detected by \fermi{};][]{abbott17b} and the late-time radio and X-ray follow-up confirming the presence of an off-axis jet \citep[e.g.][]{mooley18b,ghirlanda19,troja19gw} that strongly supported the link between BNS mergers and SGRBs.

The detection of the electromagnetic counterpart to a aLIGO\slash Virgo-detected BNS merger is of great importance as it enables the localisation of the source, along with providing complementary information such as an independent distance measurement, insight into the central engine, the energy released, and the final merger remnant. However, the initial localisation of a GW event by aLIGO/Virgo is tens to hundreds of square degrees, making it difficult to search for counterparts. 
We therefore introduce a method designed to exploit the established link between GW-detected BNS mergers and SGRBs by using the Australia Telescope Compact Array (ATCA) rapid-response mode to trigger on \swift{}-detected SGRBs. 

While the radio emission from SGRBs is usually short-lived \citep[$\lesssim2$\,days;][]{fong15}, the ATCA rapid-response mode is capable of being on-source within $10$\,minutes. 
By rapidly responding to \swift{} SGRB triggers, ATCA can become a new diagnostic tool for
uncovering the range of radio behaviour shown by SGRBs to help interpret what to look for from GW events that have off-axis gamma-ray jets. 
As targeted observations can usually reach deeper sensitivities than wide-field surveys, ATCA observations can provide
a template of the radio brightness and timing properties of BNS mergers, which will 
in-turn inform the follow-up strategies of the next era of aLIGO/Virgo GW events by wide-field radio telescopes, such as Australian instruments like the Murchison Widefield Array \citep[MWA;][]{tingay13} and the Australian Square Kilometre Array Pathfinder \citep[ASKAP;][]{johnston08}.

The jet launched during an SGRB is expected to produce a radio afterglow as predicted by the fireball model \citep{cavallo78,rees92}. 
In this model, the relativistic ejecta interact with the circumstellar medium (CSM) producing a forward shock that accelerates electrons and generates synchrotron emission. Reverse shock synchrotron emission, produced by the shock that propagates back into the post-shock ejecta, may also be observed  
depending on the density of the CSM and the ejecta. 
The broadband spectrum produced by the jet interactions in the GRB afterglow is described by the peak flux and 3 characteristic frequencies ($\nu_m$, the minimum electron energy frequency; $\nu_{sa}$, the synchrotron self-absorption frequency; and $\nu_c$, the electron cooling frequency), which evolve over time \citep{sari98,wijers99,granot99}. Only early-time radio observations are able to properly constrain 2 of these 3 frequencies ($\nu_m$ and $\nu_{sa}$), and also disentangle the reverse and forward shock components. 
By combining ATCA observations with 
multi-wavelength observations to perform SED modelling, these parameters can be derived, thus providing information about the blast wave kinetic energy, the CSM density, the magnetic field energy density and the power law electron energy distribution \citep{sari98,wijers99,granot99}. 
Limits on the linear polarisation of the reverse shock can also provide information on the jet magnetic field structure \citep{granot14}.
Early-time radio observations of SGRBs are also sensitive to temporal steepening from the jet-break \citep{sari99},
which constrains the jet opening-angle used to calculate the true energy released \citep[and therefore merger BNS/GW event rates, e.g.][]{fong14,fong15,deugartepostigo14}. 
Even early-time non-detections in the radio band can allow us to make predictions about when the forward-shock emission may peak, which can inform the cadence and duration of follow-up radio observations, potentially optimising the success of a late-time detection as we demonstrate in this paper.  
In addition, sensitive, multi-frequency, high-cadence radio observations may allow us to distinguish between more exotic emission models caused by the ejection of neutron star material or the propagation of shocks caused by the merger event, which may produce non- to ultra-relativistic omnidirectional radio emission \citep[e.g.][]{nakar11,kyutoku14}.
It is therefore crucial to obtain early-time radio observations (within minutes to days) of a larger sample of SGRBs to better characterise the timescales and frequencies necessary for understanding the range of behaviours we might expect from GW radio counterparts. 

There are also several BNS merger models that suggest 
a short-lived, supramassive and highly magnetised neutron star (NS) or ``magnetar'', supported by rotation, can exist for a short time ($<10^{4}$\,s) before finally forming a stable magnetar or further collapsing into a black hole \citep[BH, e.g.][]{usov92,zhang01,falcke14,zhang14}.  
Evidence for such merger products comes from the detection of a ``plateau phase'' in some SGRB X-ray light curves between $\sim10^{2}-10^{4}$\,s post-burst, where this departure from power-law decay indicates ongoing energy injection \citep{rowlinson13}. 
Such merger remnant scenarios may be sources of prompt, coherent radio emission \citep[see][for a review]{rowlinson19}. However, no continuous monitoring of the radio behaviour has yet been performed at GHz frequencies during the plateau phase. Such detections or upper limits could constrain different central engine models as has been done at late-times \citep[e.g.][]{fong16}. 

Only eight SGRBs have published detections in the radio band to date: GRB 050724A, 051221A, 130603B, 140903A, 141212A, 150424A, 160821B and 200522A
\citep{berger05,soderberg06,fong14,fong15,fong17,troja16,zhang17,troja19,lamb19,fong21}.
Note that this does not include GW170817 as it had a far more off-axis outflow 
than standard cosmological SGRBs so the corresponding radio afterglow was detected much later when the ejecta had moved into our line-of-sight \citep{mooley18b}. 
Out of a sample of $>70$ radio-observed SGRBs, only $\sim10$\% have been detected in the radio band at GHz frequencies \citep{fong21}.
This low detection rate may be due to an observed fast rise in radio emission with a potentially short radio afterglow lifetime. 
For example, 7 of the 8 radio-detected SGRBs were detected within 1\,day post-burst, at least half of which faded below detectability within $\sim2$ days (see Figure~\ref{fig:lc}). 
Given these short timescales, it is possible the radio emission is frequently dominated by the reverse-shock \citep[as was the case for GRB 051221A;][]{soderberg06} since simulations of BNS mergers demonstrate forward shock radio emission may evolve over days to weeks \citep{hotokezaka16} as is also the case for many LGRBs \citep[e.g.][]{vanderhorst08,vanderhorst14}. 
If we instead compare the radio-detected sample to those SGRBs that were initially observed at radio wavelengths $<1$\,day post-burst, this gives a much higher radio detection rate of $\sim30$\% \citep{fong15}. 
However, while the first four radio-detected SGRBs showed initial flux densities of $>0.1$ mJy/beam at GHz frequencies, few of the other $<1$\,day post-burst pre-2016 observations had sufficient sensitivity to detect a predicted peak flux density of $\sim40\mu$\,Jy/beam at 10\,GHz for an SGRB at an average redshfit of $z=0.5$ with an expected CSM density of $n_{0} \sim0.1$\,cm$^{-3}$ \citep{berger14}. In fact, the four most recent radio-detected SGRBs peak at $\lesssim40\mu$\,Jy/beam.

The small sample of radio detected SGRBs therefore provides limited knowledge of their radio afterglow brightnesses and timescales, and is insufficient for deriving the energy outputs and environmental properties of the population through multi-wavelength modelling. It is therefore vital to perform both rapid and sensitive radio follow-up observations of SGRBs to capture these short-lived and faint events. The key to achieving this is through the use of rapid-response (also known as triggering) systems, where a telescope has the ability to automatically respond to a transient alert, and either repoint at the event or update its observing schedule to begin observations when the source has risen above the horizon. Rapid-response radio telescopes have been in use since the 1990's \citep[for example see][]{green95,dessenne96,bannister12,palaniswamy14,kaplan15} 
but predominantly at low radio frequencies (100\,MHz to 2.3\,GHz), with the majority of experiments being designed to search for prompt, coherent radio emission. However, until recently, the only high frequency ($>5$\,GHz) rapid-response program designed to target incoherent (synchrotron) radio emission from GRBs has been run on the Arcminute Microkelvin Imager (AMI) Large Array (LA), known as ALARRM (the AMI-LA Rapid Response Mode), which has been active since 2012 \citep[][]{staley13,anderson18}. It was only through ALARRM that it was possible to be on-source fast enough to detect the rise and peak in the reverse-shock radio emission at 15\,GHz from GRB 130427A within 1\,day post-burst, which also represents one of the earliest radio detections of a GRB to date \citep{anderson14}. In addition, the radio catalogue of AMI observations of 139 GRBs (12 were short GRBs but non-detections), the majority of which were automatically triggered on using the rapid-response mode within $1$\,day post-burst, was the first representative sample of GRB radio properties that was unbiased by multi-wavelength selection criteria \citep{anderson18}. This work revealed that possibly up to $\sim44-56$\% of \swift{}-detected LGRBs have a radio counterpart (down to $\sim0.1-0.15$\,mJy/beam), with the increase in detection rate from previous studies \citep[$\sim30$\%;][]{chandra12} likely being due to the AMI rapid-response mode, which allows observations to begin while the reverse-shock is contributing to the radio afterglow. This program has motivated the installation of a rapid-response mode on the ATCA. 

Here we present the first triggered observation of a SGRB using the new ATCA rapid-response mode. 
ATCA is an ideal instrument for performing triggered radio follow-up of \swift{} SGRBs due to its high sensitivity and broadband receivers that provide simultaneous multi-frequency coverage. The ATCA response times (which can be as short as minutes) 
have the potential to be 
much faster than the current median SGRB response of the Karl G. Jansky Very Large Array (VLA; $\sim24.7$\,hrs), which rely on manually scheduling target-of-opportunity observations \citep{fong15}.
In Section 2, we describe the ATCA rapid-response system from the observer interaction (front-end) level and the observatory (back-end) level.
In Section 3, we describe the triggered ATCA observation and data reduction of GRB 181123B, and corresponding results. 
This is followed by a comparison of our radio limits for GRB 181123B to the sample of radio-detected SGRBs and a discussion of the parameter space that the triggered ATCA observations are probing in Section 4. 
Finally, we perform modelling of the GRB 181123B afterglow 
and thus demonstrate the usefulness of obtaining early-time (within 1 day) radio observations of an SGRB (regardless of whether or not there is a detection) to place constraints on the GRB physics.  

\section{ATCA rapid-response mode}

ATCA is a six element, 22\,m dish, East-West interferometer based in New South Wales in Australia. Its maximum baseline length is 6\,km and it is capable of observing in multiple, broad frequency bands  
with full polarisation, and in a variety of array configurations. 
ATCA is currently equipped with the Compact Array Broadband Backend \citep[CABB;][]{wilson11}, which has a 2\,GHz bandwidth that is capable of observing in two frequency bands simultaneously with tunable receivers that operate between 1.1-105\,GHz. 

Since 2017 April 18, ATCA has been capable of rapidly responding to transient alerts. 
The rapid-response mode can trigger using the 16\,cm, 4\,cm and 15\,mm receivers, corresponding to a usable frequency range of $1.1-25$\,GHz, and can observe in any CABB mode. 
In the following, we describe both the observer front-end and the observatory back-end of this new triggering system.

\subsection{VOEvent parsing/front-end}\label{sec:front-end}

The front-end software we use to interface with the ATCA rapid-response system ({\sc vo\_atca})\footnote{https://github.com/mebell/vo\_atca} is designed to trigger on Virtual Observatory Events (VOEvent; \citealt{seaman11}), which are the standard format for broadcasting machine readable astronomical alerts related to transient events. 
A VOEvent package contains all the required data (in {\sc xml} format) that allow automated decisions to be made in real-time given certain keywords and parameters. 
VOEvents are brokered via the 4 Pi Sky VOEvent Broker \citep{staley16pp} and the {\sc comet} VOEvent client \citep{swinbank14}. 
These packages allow us to listen to multiple VOEvent streams, including those broadcast by \swift{}. 
We use the {\sc Python} package {\sc vovent-parse} \citep{staley_voevent-parse_2014} as the main tool to read the VOEvents 
and to extract the required information to be assessed by the triggering algorithm. 

Upon receiving a \swift{} VOEvent, the ATCA VOEvent parser uses the keyword {\sc grb\_identified = true} to initially identify a GRB packet. 
Packets containing {\sc startrack\_lost\_lock=true} are ignored as it means that \swift{} has lost its celestial position-lock so such an alert is unlikely to be from a real transient.  
While the observatory back-end prevents the telescope from overriding for sources that are too far north (see Section~\ref{sec:back-end}), we impose an additional declination cut-off for all SGRBs north of +15$^{\circ}$ to 
ensure the potential for $>8$\,hr integrations for the triggered observations. 

On passing these stages, the parser then assesses the duration of the trigger so that SGRB candidates can be identified. However, on the short timescales following the alert, and with growing uncertainty as the GRB burst duration increases, it is difficult to classify \swift{} GRBs as short or long in an automated way.
A rigorous classification of the GRB requires human inspection of the data, which is only published online on the Gamma-ray Coordinates Network Circulars (GCN)  Archive,\footnote{https://gcn.gsfc.nasa.gov/gcn3\_archive.html} usually between 10\,mins and 1\,hr post-burst and therefore not via a VOEvent. 
To account for this, we implemented a three-tiered system to flexibly respond to different GRB durations and therefore filter for those events more likely to be SGRBs. 
The keyword {\sc integ\_time} (the length of time for the transient signal to reach a significant threshold) parameter is used as an estimator of the incoming GRB's true duration. 

\begin{itemize}
\item GRBs with {\sc integ\_time}$<$0.257\,s have a high probability of being SGRBs so the VOEvent parser will automatically submit these triggers to the observatory and alert team members via text and email of the override observation. 

\item With durations $0.257$\,s\,$<${\sc integ\_time}\,$<1.025$\,s, we have implemented a "wait-to-proceed" algorithm as the probability of the GRB being a SGRB decreases with increasing {\sc integ\_time}. In this case, we issue email and text alerts so that team members can check the GCN Archive for adequate verification of the GRB classification. If the GRB is confirmed to be short, then the duty team member responds "YES" to the detection email, and this email reply is read by an algorithm (via the Google email Application Programming Interface\footnote{https://developers.google.com/gmail/api}) 
%API) 
that then proceeds with submitting the trigger to ATCA, resulting in an override observation. This provides an easy interface to assess and submit triggers via a mobile phone, which can receive SMS alerts and allow responding to emails away from a computer. 

\item If {\sc integ\_time}$>$1.025\,s then we presume that the GRB is long and we do not proceed with submitting a trigger to override the telescope. 
\end{itemize}

After the parser (or duty team member) has successfully identified the event as an SGRB, 
our algorithm then searches the ATCA calibrator database for a nearby and suitable phase calibrator.  
It then automatically builds a schedule file (we use the ATCA scheduler software {\sc cabb-schedule-api})\footnote{https://github.com/ste616/cabb-schedule-api} for a 12-hour observation of the GRB in the requested frequency band (for GRB triggering we currently use the 4\,cm receiver), which has interleaved 
phase calibrator observations every 20 minutes. 
Note that the total exposure time is also limited by how far the GRB is above the horizon at the time of the trigger. 
The schedule file and override request is then submitted to the observatory where it is assessed for submission to the observing queue by the ATCA back-end.

\subsection{Observatory back-end}\label{sec:back-end}

Time on the ATCA is scheduled into two 6-month long semesters, and the order of observations in each semester is set months in advance. This is done to allow the project investigators, who are also responsible for conducting the observations, to plan their activities. A rapid-response system is not easily compatible with this mode of operation.

Nevertheless, demand for the telescope to quickly respond to events has been steadily rising. In 2016, roughly 10\% of telescope time was given to  
NAPA (Non A-priori Assignable) or ToO (Target of Opportunity) projects, while in 2019 this figure had risen to 19\%. 
For a NAPA project, a science case is given to the time assignment committee (TAC), which ranks its significance against the other projects for that semester. Provided the science is considered compelling, these projects are allowed to displace time from other projects during the semester, with the philosophy being that were we to know during the scheduling process when an event would happen, a compelling project would have been scheduled to observe it.

Rapid-response NAPAs operate in the same way. A scientific justification must be supplied to the TAC, who must agree that rapid response is warranted. The observatory then supplies an authentication Javascript Web Token (JWT) to the project, and assists the investigators to test their automatic triggering system.

A web service is provided so that the trigger to start observations can be sent from any internet-connected device. A Python library ({\sc atca-rapid-resonse-api}) is also available to make it easier to send requests to this service.\footnote{https://github.com/ste616/atca-rapid-response-api} All requests must contain a valid schedule file, and must nominate the target coordinates and a nearby phase calibrator.

Upon receipt of a trigger, the web service tries to schedule the observation as soon as possible. If the source is above the horizon and the user-nominated minimum useful observing time can be obtained before the source sets, the current and subsequent observations can be displaced  
and the system can start the observations within 2 seconds of the trigger's arrival. Within that time, emails are sent to the projects that will be displaced, and to the triggering team, describing the new order of the scheduling. The schedule is also altered as necessary to add a scan on a flux density calibrator at an opportune time, and potentially to shorten the observations to fit the available time. At all times, the emphasis is to move the telescope to the target coordinates as quickly as possible.

The service can also provide an immediate rejection should no suitable time be found for the observation. For example, if no available time can be found up to 100 hours in the future (generally because the request was made during a time when the array is shutdown for maintenance or participating in VLBI observing), the observations are rejected and the proposal team are notified. 
While no explicit limit is set for the source declination, sources too far north may not be available for the user-nominated minimum useful observing time, and will thus be rejected.

If the web service can schedule the observations, a separate service then takes over, and takes control of the observing control software. Some more checks are made to see if the array can be used for observing, and will delay the start of the observations if the weather conditions are unsuitable. This service also monitors the observations for interruptions due to weather, equipment failure and human intervention. Rudimentary responses are pre-programmed for any such eventuality. The service stops once the observations have finished, the target sets, or the observations are cancelled, whichever comes first. Control of the telescope then goes to the investigators whose project was scheduled to be running at this end time.

A more complete guide to the operation of the rapid-response system is provided in the ATCA Users Guide.\footnote{https://www.narrabri.atnf.csiro.au/observing/users\_guide/html/atug.html}

\subsection{Triggering performance} \label{sec:trig_per}

Since the commencement of the program, we have worked with the observatory to improve the success of SGRB triggered observations with ATCA, which involved extensive system and software debugging. Many SGRBs were missed due to the telescope being in uninterruptible modes such as maintenance, reconfiguration, participating in VLBI or operating in an incompatible correlator mode (the latter has since been resolved). 

Our original override strategy involved triggering on all \swift{} GRBs with {\sc integ\_time}$<1.025$\,s as SGRBs have been detected with {\sc integ\_time} up to $1.024$\,s. However, as mentioned in Section~\ref{sec:front-end}, the majority of events within this timescale are LGRBs. 
\swift{} data requires a human in the loop to classify the event as long or short, which is usually based on the duration and the hardness of the event (note that SGRBs often produce higher energy prompt emission than LGRBs) and are only published on the GCN Archive up to an hour post-burst \citep[also note that the distinction between events with durations between $1-2$\,s can be tenuous and has led to discussions regarding intermediate GRB classes; e.g.][]{mukherjee98,horvath98,huja09,deugartepostigo11}. 
This original strategy therefore resulted in several false ATCA triggers, most of which were identified and cancelled before the telescope was overridden as there was additional lead time before the event in question had risen above the horizon. However, 
there were a few instances where some data was collected on LGRBs. 
Recent edits to the VOEvent parsing of event timescales using the keyword {\sc integ\_time}, which are described in Section~\ref{sec:front-end} have resulted in a significant reduction in ATCA triggers of LGRB contaminants. 

When ATCA receives a trigger of an event that is above the horizon, the main limitation to the response time is the telescope slew speed. On receiving the VOEvent via the parsing code, it takes $2-3$\,s for the observation to be queued and the subsequent maximum observing time calculated. Following a \swift{} alert on the long GRB 190519A \citep{ukwatta19}, ATCA was on target and observing the event in 2\,min and 39\,s. Other response times range between $3-6$\,min post-burst, which make the ATCA rapid-response system competitive with other triggering facilities such as AMI \citep[e.g.][]{anderson18} yet is also based in the Southern Hemisphere, has more collecting area, a larger number of frequency bands, polarisation capabilities, and (in some configurations) better angular resolution.

\subsection{Short GRB experimental design}

The majority of GRBs detected by \swift-BAT are LGRBs, with SGRBs (in this case events with $T_{90} \leq 2$\,s including those found in ground analysis) only accounting for $\sim7-8$\% 
(this is based on event numbers between 2017 and 2019 using the \swift{} GRB Look-up Table,\footnote{https://swift.gsfc.nasa.gov/archive/grb\_table/} where $T_{90}$ is the time between 5 and 95\% of the fluence being emitted). We therefore expect $\sim5-10$ SGRBs to be detected by \swift{} per year, and therefore predict $\lesssim2$ will be observable with ATCA (below a Declination cut-off of +15\,deg) during an observing semester. 

Our rapid-response observations are performed using the $4$\,cm receiver, which has dual 2\,GHz windows that are usually centered at 5.5 and 9\,GHz, which is the most sensitive ATCA band. This choice is based on several factors: the full-width half-maximum of the primary beam encompasses the initial \swift{}-BAT positional uncertainty of newly detected GRBs \citep[1-4 arcmin;][]{barthelmy05}, it is largely immune to atmospheric instabilities, and is less disrupted by RFI than other ATCA bands.
In addition, as synchrotron emission from GRB reverse and forward shocks peaks earlier and brighter with increasing frequency, the 4\,cm band (5.5/9 GHz) is optimal for ensuring the source will be bright but not peaking before the telescope is on-target.

As mentioned in Section~\ref{sec:intro}, the radio afterglows from SGRBs are usually detected within 1\,day post-burst \citep[e.g.][]{fong15}, which strongly motivates our need for the ATCA rapid-response mode. The triggered observations are designed to  observe between $2-12$\,h (depending on how long the source is above the horizon following the trigger). 
As previous SGRB radio studies have shown that the radio afterglow has already switched-on within $4-16$\,h post-burst \citep[e.g.][]{anderson18}, a $\leq12$\,hr observation allows us to track the rapid rise in emission with a sensitivity of $\sim60 \, \mu$Jy ($3\sigma$) on one hour timescales.\footnote{https://www.narrabri.atnf.csiro.au/myatca/interactive\_senscalc.html} This means that any delays of $\leq1$\,hr related to waiting for the GRB classification does not affect the rapid-response science goal (see Section~\ref{sec:front-end}). 
A $\leq12$\,hr track also ensures some periods of simultaneous \swift{} X-ray Telescope \citep[XRT, observing band between $0.3-10$\,keV;][]{lien18} observations, 
which is essential for modelling the spectral energy distribution (SED), and for exploring the radio properties associated with the plateau phase (e.g. see our modelling in Section~\ref{sec:mod}). 

\begin{figure*}
  \begin{center}
  \includegraphics[width=0.49\textwidth]{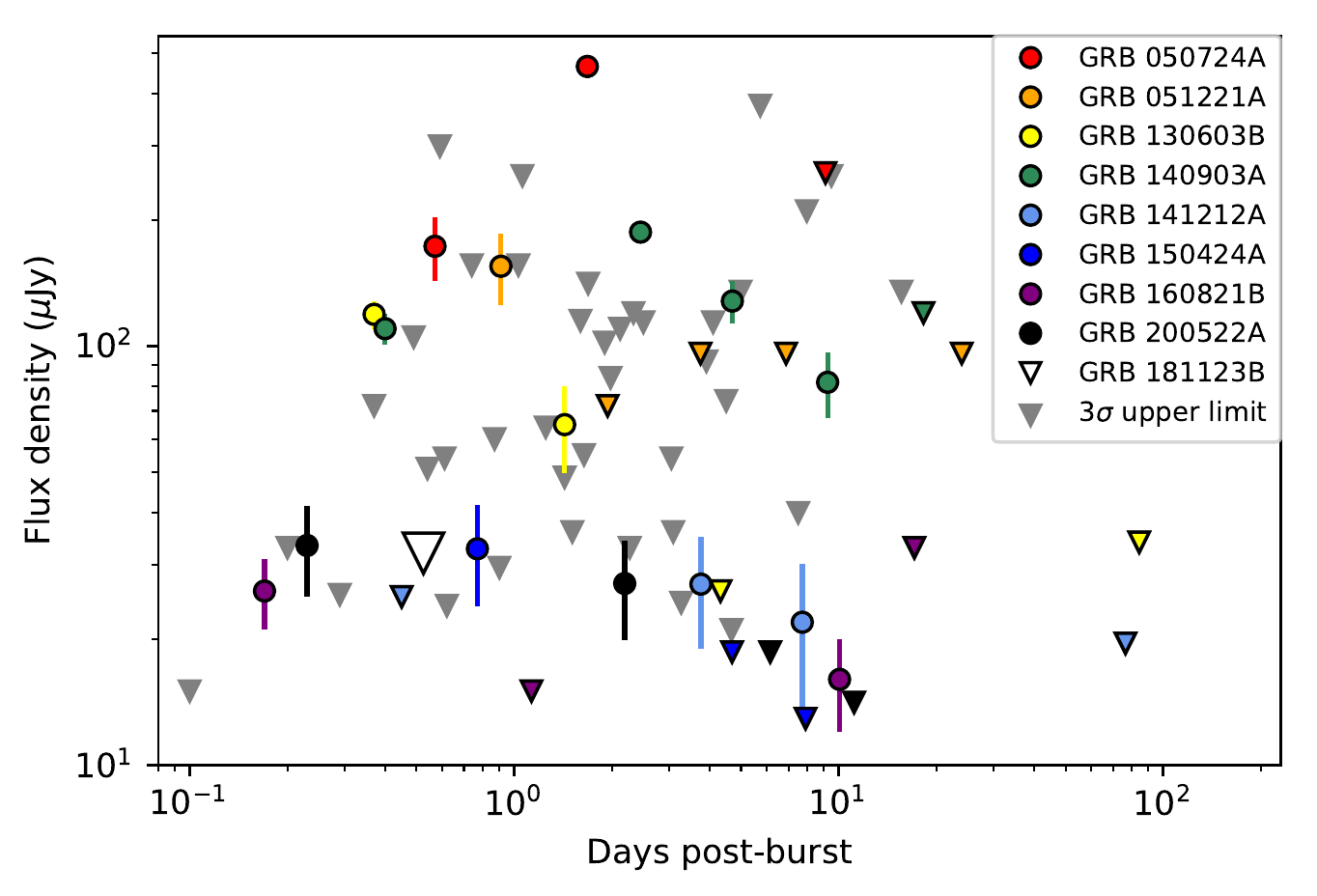}\label{fig:flux}
  \includegraphics[width=0.49\textwidth]{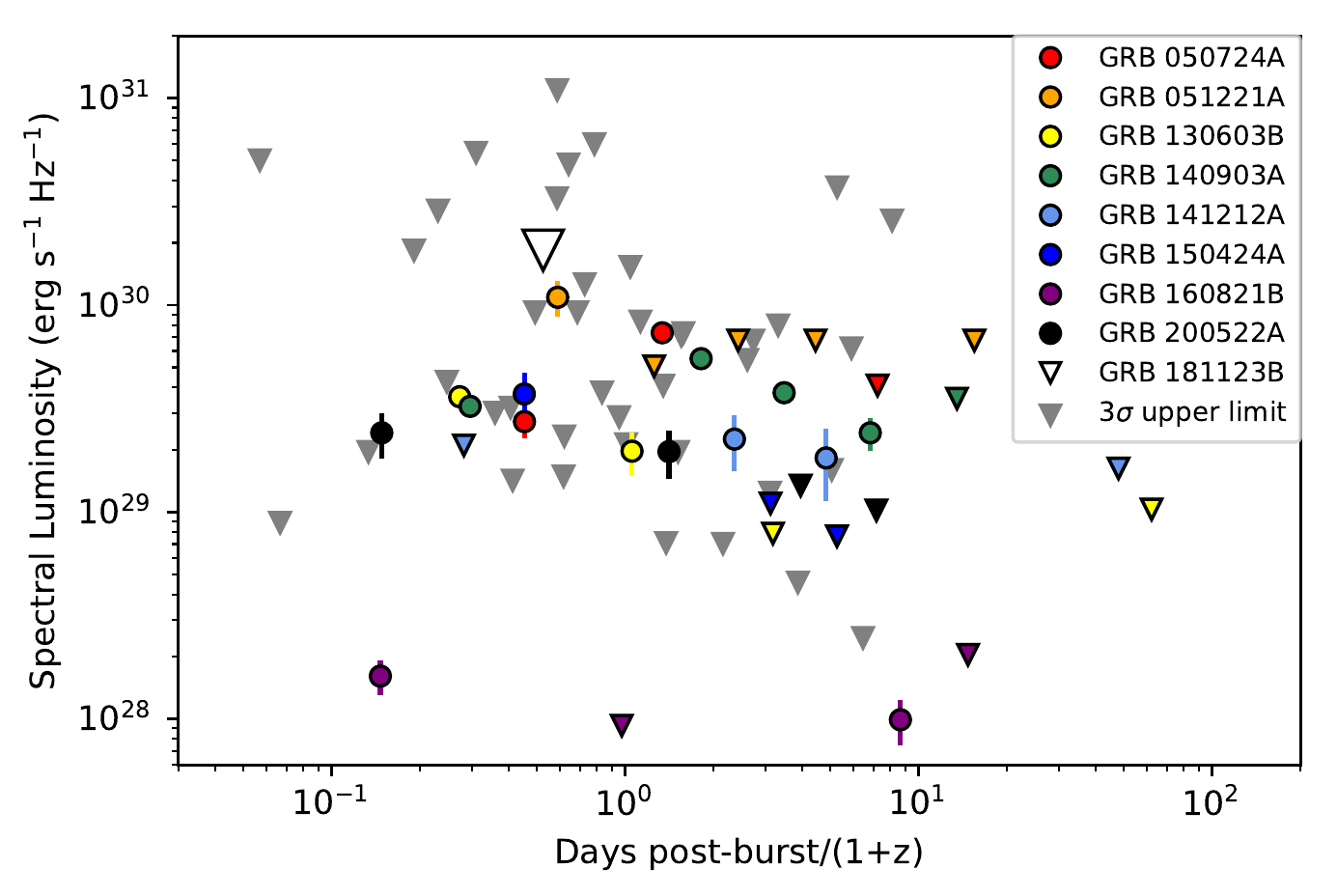}\label{fig:lum}
  \caption{Radio light curves of SGRB radio detections ($1\sigma$ error bars) and $3\sigma$ upper-limits observed at frequencies between 6 and 10\,GHz. 
  Left: Radio flux density vs days post-burst and Right: $k$-corrected spectral luminosity vs days post-burst in the rest-frame. The 9\,GHz upper-limit of GRB 181123B is depicted as a large white triangle. For those GRBs without a known redshift we assume $z=0.5$. The $3\sigma$ upper limits of those SGRBs that were observed but not detected in the radio band are depicted as grey triangles.  
  References for all radio flux densities and redshifts for radio-detected SGRBs: \citet{berger05}, \citet{fox05}, \citet{prochaska05}, \citet{soderberg06}, \citet{cucchiara13}, \citet{deugartepostigo14}, \citet{cucchiara14}, \citet{chornock14}, \citet{fong14}, \citet{fong15}, \citet{troja19}, \citet{lamb19}, \citet{paterson20}, \citet{fong21}. All radio upper limits shown in grey were taken from \citet[][see references therein]{fong15}.}
  \label{fig:lc}
  \end{center}
\end{figure*}

Following the triggered, rapid-response observation, we also request three $\sim4$\,hr follow-up observations in the $4$\,cm band to occur between $1-3$, $4-6$, and $8-12$\,days post-burst, which can reach a sensitivity of $30\,\mu$Jy ($3\sigma$). 
While 3 of the previous radio-detected SGRBs faded below detectability within 2 days post-burst, the other 2 were detected up to 10\,days post-burst (see Figure~\ref{fig:lc}), thus motivating this more long-term monitoring of any triggered candidate. 

\section{ATCA observations of GRB 181123B}

\swift{}-BAT detected the short GRB 181123B at 05:33:03 UT (trigger=873186), which was rapidly detected in the X-rays by the \swift{}-XRT  
and localised to the position $\alpha \mathrm{(J2000.0)} =  12^{\mathrm{h}}17^{\mathrm{m}}28\overset{\mathrm{s}}{.}05$ and $\delta (\mathrm{J2000.0}) = +14^{\circ}35'52\overset{''}{.}4~$ with a 90\% confidence of $1\overset{''}{.}8$ \citep{osborne18}. Further optical and near-infrared follow-up detected a source coincident with the \swift{}-XRT position \citep{fong18,paterson18a,paterson18b}, 
resulting in the identification of the host galaxy at redshift $z=1.754$ and the detection of the optical afterglow to GRB 181123B \citep[$i=25.1$\,mag at $9.1$\,h post-burst;][]{paterson20}. This makes GRB 181123B one of only three SGRBs at $z>1.5$ \citep{paterson20}. 

On receiving the VOEvent trigger, ATCA was automatically scheduled to begin observations on 2018 Nov 23 at 18:07:24.9 UT (12.6\,h post-burst) for 8.3\,h  \citep{anderson18gcn}, when the GRB had risen above the horizon (minimum elevation of 12\,deg). On this date, ATCA was in the 6B array configuration, and the triggered observations were taken in the 4\,cm band, with the dual 2\,GHz bandwidth windows centered at 5.5 and 9\,GHz. The observation pointing was at the initial BAT position, which was $1.2$\,arcmin offset from the final \swift{}-XRT position of GRB 181123B. 
Note that we requested no follow-up ATCA observations due to the imminent reconfiguration and correlator reprogramming, with many subsequent programmes having priority.

The ATCA rapid-response observation was reduced and analysed with the radio reduction software {\sc MIRIAD} \citep{sault95} using standard techniques. Flux and bandpass calibration were conducted using PKS 1934-638 and phase calibration with PKS 1222+216. Several rounds of phase and amplitude self calibration were also applied \citep[this was possible due to the nearby bright field source FIRST~J121731.7+143953;][]{helfand15}. In order to obtain the most robust flux density upper limits at the position of the GRB, we used {\sc mfclean} to create a clean model of the sources in the field (manually drawing clean boxes) and subtracted this model from the visibilities. A primary beam correction was then applied due to the 1.2\,arcmin offset between the pointing centre and the best known GRB position from the \swift{}-XRT. GRB 181123B was not detected, and the final $3\sigma$ upper-limits can be found in Table~\ref{tab:obs}. 

As we know the precise location of GRB 181123B to within the ATCA beam, we also report the peak force-fitted flux density at both 5.5 and 9\,GHz in Table~\ref{tab:obs}.
These were calculated using the task {\sc imfit} to force-fit a Gaussian to the beam that was fixed at the \swift{}-XRT position of the GRB (errors are the $1\sigma$ rms).
The advantage of quoting the force-fitted flux density over an upper-limit is that such a measurement also accounts for the presence of nearby sources, as well as variations in the noise across the image.
The data were also divided into 3\,h and 1\,h timescales and then re-imaged to search for evidence of emission that may have switched on nearer the end of the observation; however none was detected. 

\begin{center}
\begin{table}
\caption{ATCA observations of GRB 181123B at 5.5 and 9 GHz, which began on 2018 Nov 23 at 18:07:24.9 UT (12.6\,h post-burst) for 8.3\,h.}
\label{tab:obs}
\begin{tabular}{lcc}
\\
\hline
Frequency & $3\sigma$ Upper-limit & Forced-fit flux density\\
(GHz) & ($\mu$Jy/beam) & ($\mu$Jy/beam)\\
\hline
 5.5 & 34 & $7 \pm 12$ \\
 9.0 & 32 & $15 \pm 11$ \\
\hline
\end{tabular}
\end{table}
\end{center}

\section{Discussion}

In this section, we first demonstrate that our radio flux density limits for GRB 181123B are consistent and competitive with previous studies of the radio-detected SGRB population. This is followed by afterglow modelling to demonstrate the importance of obtaining early-time radio observations (regardless of whether there is a detection) to better constrain the properties of the blast wave.

In Figure~\ref{fig:lc}, we show the light curves of SGRBs observed in the radio band between 6 and 10\,GHz. The 8 radio-detected SGRBs are colour-coded with $3\sigma$ upper-limits represented by triangles. The $3\sigma$ upper limits of those SGRBs observed but not detected in the radio band have been plotted as grey triangles. The ATCA 9\,GHz $3\sigma$ upper-limit of GRB 181123B is shown as a large white triangle. In the left panel of Figure~\ref{fig:lc}, we have plotted the observed radio flux density vs days post-burst, whereas in the right panel we have plotted the spectral luminosity vs days post-burst in the rest frame, assuming a redshift of $z=0.5$ \citep{berger14} for those events with no known redshift.  
When converting the flux ($F$) to luminosity ($L$), a $k$-correction was also applied such that $L=4 \pi F d_{L}^{2} (1+z)^{\alpha - \beta -1}$\,erg\,s$^{-1}$\,Hz$^{-1}$, where $d_{L}$ is the luminosity distance for the redshift $z$ \citep[assuming $\Lambda$CDM cosmology with $H_{0}=68$\,km\,s$^{-1}$\,Mpc$^{-1}$ and $\Omega_m=0.3$;][]{planck16}, and $\alpha$ and $\beta$ are the temporal and spectral indices defined as $F \propto t^{\alpha}\,\nu^{\beta}$ \citep{bloom01}. We assume $\alpha=0$ and $\beta=1/3$, which are appropriate for an optically thin, post-jet-break light curve \citep[see][]{chandra12}.

From Figure~\ref{fig:lc} we can see that the ATCA flux limit for GRB 181123B is extremely competitive and consistent with the most constraining lower-limits. 
Using formalism by \citet{granot02}, \citet{berger14} showed that if we assume fiducial parameters for SGRBs, along with typical microphysical parameters for LGRBs, the expected peak flux density at a redshift of $z=0.5$ is $F_{\nu}\sim40\mu$\,Jy at $\sim10$\,GHz for an ambient medium density of $n_{0}=0.1$\,cm$^{-3}$. 
Our $3\sigma$ sensitivity at 9\,GHz was 32$\mu$\,Jy, and therefore sensitive enough to detect emission from a GRB with the above properties, however, it is important to note that some GRB microphysical and macrophysical parameters like the kinetic energy and the CSM density can vary by several orders of magnitude \citep{granot14}.

The luminosity light curves in Figure~\ref{fig:lc} show the $3\sigma$ upper-limit for GRB 181123B at $\sim1$\,day post-burst (in the rest frame). 
Given the very high redshift of GRB 181123B, even these sensitive ATCA observations would not have detected the radio counterpart to the seven SGRBs detected at early times (within a day post-burst in the rest frame) if they were placed at $z=1.754$.
We therefore cannot draw any further comparisons between the physical properties of the radio-detected GRB sample and GRB 181123B based on luminosity alone and require more detailed multi-wavelength light curve modelling (see Section~\ref{sec:mod}).

\subsection{Modelling constraints} \label{sec:mod}

In this section, we model the afterglow of GRB 181123B in order to explore how early-time ($<1$\,day) radio observations of SGRBs can help to constrain the dynamical and microphysical parameters of such blast waves in the context of the fireball model. 
Using the redshift derived from the identification of the host galaxy of GRB 181123B \citep[$z=1.754$;][]{paterson20}, 
we model the force-fitted flux density values at the \swift{}-XRT position of the GRB from our ATCA observations together with the \swift-XRT light curve \citep{evans09,evans10}. 
We have chosen to use the force-fitted flux measurements plus errors in our modelling as it allows us to assign a likelihood to a predicted model flux for a set of model parameters, which is not possible with an upper limit 
\citep[for some examples of where radio force-fitted flux measurements are quoted and used in afterglow modelling see][]{galamawijers98,kulkarni99,vanderhorst11,vanderhorst15}. 

For this modelling, we have chosen to only consider the forward-shock component to minimise complexity, particularly as we are dealing with a small number of data points. 
As previously mentioned, 
the reverse-shock could be dominant at early times ($\lesssim1$\,day) in the radio band as has been observed for some SGRBs \citep[e.g.][]{soderberg06,lamb19,troja19}. 
Given that the reverse-shock evolves to lower frequencies more rapidly than the forward-shock and we have no radio detection, our modelling depends primarily on the X-ray detections, which are always dominated by the forward-shock, thus motivating our model choice. 
Our afterglow fitting also does not rule out a reverse shock contribution.  
We therefore assume a spherical, relativistic, blast wave interacting with the circumburst medium and generating synchrotron emission. Since SGRBs are known to occur in homogeneous, low density environments \citep[median densities of $n_{0}\approx (3-15) \times 10^{-3}$\,cm$^{-3}$ with $\approx80-95\%$ of events being situated in environments of $n_{0}<1$\,cm$^{-3}$;][]{fong15}, 
we assume a constant density circumburst medium. 

We use the \texttt{boxfit} code to model the afterglow emission \citep{vanEerten2012}. \texttt{boxfit} makes use of pre-calculated hydrodynamics data to calculate the dynamics of the blast wave, and solves radiative transfer equations on the go. 
Since in this work we assume a spherical blast wave, we fix the opening angle ($\theta_0$) to $\pi / 2$. 
We then use the C++ implementation of the \texttt{MultiNest} nested sampling algorithm, which is a Bayesian inference tool, to determine the posterior distributions of the free parameters \citep{feroz09}.  The free parameters of our model are defined as:

\begin{itemize}
  \item $E_{K, \mathrm{iso}}$: Isotropic equivalent kinetic energy in units of erg.
  \item $n_0$: Circumburst medium number density in units of $\mathrm{cm}^{-3}$.
  \item $p$: Power-law index of the accelerated electron distribution, such that $N(\gamma) \propto \gamma^{-p}$, with some minimum Lorentz factor $\gamma_{m}$ \citep{wijers99}.
  \item $\epsilon_B$: Fraction of thermal energy in the magnetic fields.
  \item $\epsilon_e$: Fraction of thermal energy in the electrons.
\end{itemize}

In order to demonstrate how the inclusion of early-time radio data helps to further constrain 
the dynamical and microphysical parameters (when combined with \swift{}-XRT observations and regardless of whether or not there is a radio detection), we model the afterglow of GRB 181123B with and without the ATCA force-fitted fluxes and compare the posterior distributions of the free parameters. 
In both fits, we use the same prior for the free parameters (Table~\ref{tab:prior}) and the resulting best fit values in Table~\ref{tab:post_param} are set with the lowest chi-squared value in the posterior. 

\begin{figure}
  \includegraphics[width=.45\textwidth]{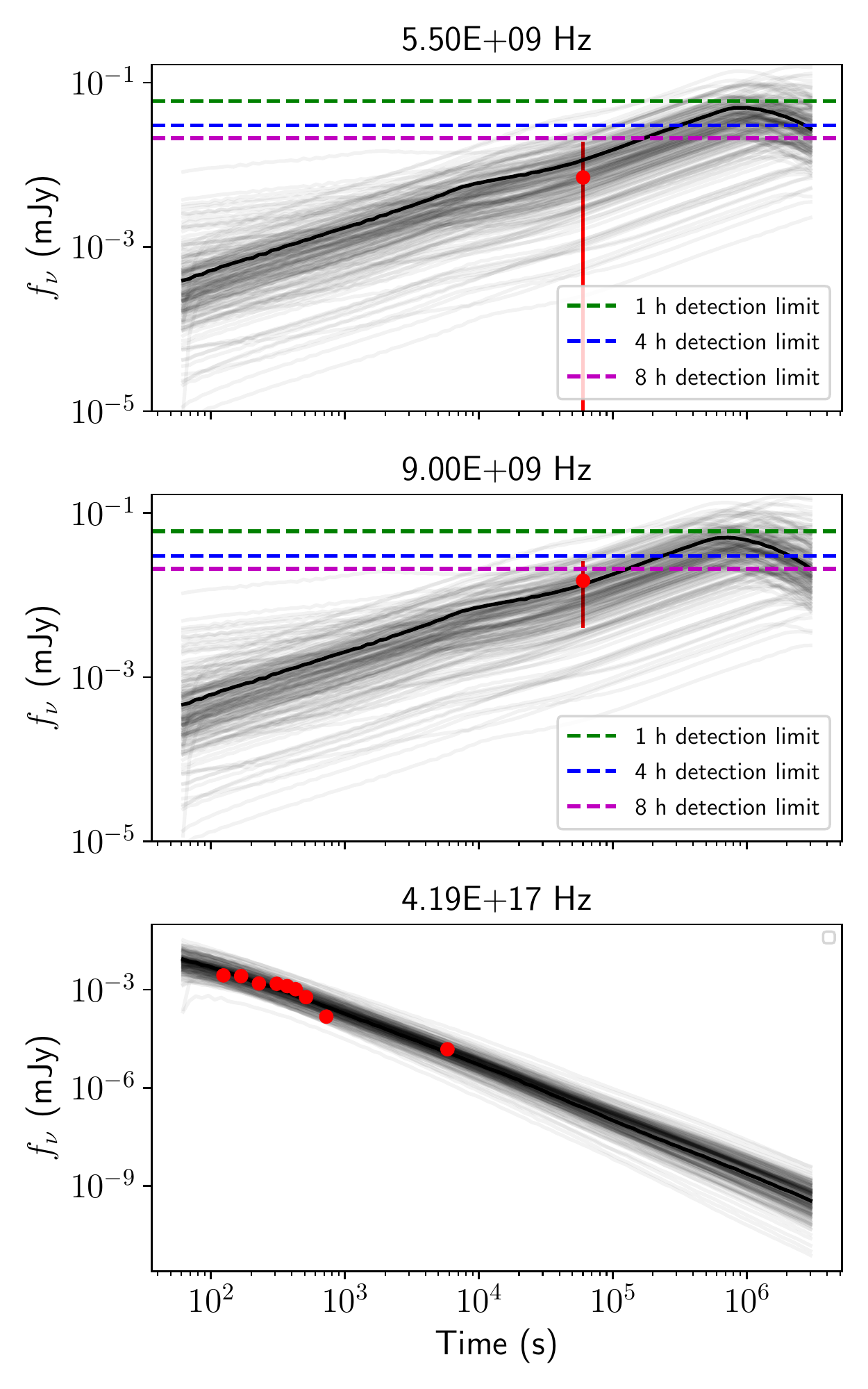}
  \caption{Fit result for the afterglow light curves of GRB 181123B at ATCA observing frequencies 5.5\,GHz (top panel) and 9.0\,GHz (middle panel), and with the \swift{}-XRT ($0.3-10$\,keV; bottom panel) when the ATCA data are included (in this case the force-fitted flux densities; red data point and error bar plotted on the 5.5 and 9.0\,GHz light curves). The plotted values in the \swift{}-XRT light curve (also red data points) were downloaded via the \swift{} Burst Analyser \citep{evans10}. For each of the three frequency bands, 200 light curves are drawn by sampling the inferred posterior distribution of the parameters. The solid line represents the best fit model. The horizontal dashed lines show the 3-$\sigma$ detection limit for various ATCA integration times.
  }
  \label{fig:lc_post}
\end{figure}

Light curves for the posterior predictive distribution when the ATCA force-fitted flux density values are included in the modelling, together with the best fit, can be found in Figure \ref{fig:lc_post}. 
Given that the modelling of the X-ray detections of GRB 181123B alone suggests an energetic solution, the inclusion of radio information aids to pull down the overall fit so that at both 5.5 and 9\,GHz, the best fit light curves are clustered around the ATCA force-fitted flux densities. 
While the resulting model is consistent with the \swift{} Ultraviolet/Optical Telescope \citep[UVOT;][]{roming05} upper-limits \citep{oates18}, it over-predicts the Galactic extinction corrected $i$-band flux reported by \citep{paterson20} by a factor of $\sim3$ or 1.2 magnitudes. 
At this high redshift, an $i$-band detection indicates the afterglow emission was produced at ultraviolet wavelengths in the rest frame, and would therefore be quite prone to extinction by dust.
Given our model does not consider extinction, intrinsic or otherwise, this over-prediction may therefore not be unreasonable. 
However, our $i$-band prediction is much higher than the host optical extinction calculated by \citet{paterson20} from photometric observations ($A_{V}$=0.23) or calculated from their observed excess hydrogen column density ($N_{H}$; derived from X-ray afterglow spectral modelling), which is known to scale to optical extinction \citep{guver09}, predicting $A_{V}$=0.38.
There are also other potential sources of optical and infrared emission from SGRBs such as a kilonova from r-process radiative decay \citep[e.g.][]{metzger10}, which our model does not include. However, such emission usually does not dominate over the afterglow until $>1$\,d post-burst \citep[e.g.][]{tanvir13}. 

As can be seen in Table~\ref{tab:post_param}, the inclusion of the ATCA force-fitted fluxes in our modelling allows for much better constraints to be placed on $\epsilon_e$ (see also Figure~\ref{fig:marge_comp}, which shows a comparison between the marginal distributions of the parameters for both cases - modelling with and without the ATCA data). 
The rest of the parameters are consistent between both modelling experiments but the $E_{K,\mathrm{iso}}$ is on the brighter end of known SGRBs \citep{fong15}. 
Our findings are consistent with those by \cite{beniamini17}, who have shown that the flux density and time of the GRB radio light curve peak can be used to particularly constrain $\epsilon_e$. We also note that our constraint on $\epsilon_e$ is also consistent (within the 95\% credible interval) with the distribution of $\epsilon_e$ ($0.13-0.15$) found through the analysis of 36 GRB radio afterglows performed by \citet{beniamini17}.
The predicted radio peak also suggests that at later times ($\gtrsim3-4$\,days post-burst), the forward shock radio emission from GRB 181123B may have been detectable at 5.5 and 9\,GHz with $\geq4$\,hr ATCA integrations (see Figure~\ref{fig:lc_post}). 
Therefore, the inclusion of early-time radio data in GRB afterglow modelling (regardless of whether or not there is a detection), 
together with an X-ray light curve, allows us to predict the forward shock peak radio flux density, 
thus constraining the fraction of shock energy in the relativistic electrons  ($\epsilon_e$). 
\citet{paterson20} also derived these same afterglow parameters for GRB 181123B but assumed fixed values of $\epsilon_e=0.1$ and $\epsilon_B=0.1$ or 0.01. While our parameters are far less constrained, our values for $E_{K,iso}$ and $n_{0}$ (as well as $\epsilon_e$ and $\epsilon_B$)  are consistent with \citet{paterson20} within the 95\% credible intervals. However, our value range for $p$ was higher and did not overlap with the range derived by \citet{paterson20}. Note that our value range for $p$ is more consistent with those calculated for radio-detected SGRBs (see Section~\ref{sec:comp}).

\begin{table}
    \setlength{\tabcolsep}{5pt}
    \def\arraystretch{1.25}
    \caption{Assumed priors for the free parameters for all modelling efforts.}
    \begin{tabular}{l|r}
    \hline
    \multicolumn{1}{l}{Parameter range} & \multicolumn{1}{r}{Prior distribution} \\
    \hline
    $10^{49}<E_{K, \mathrm{iso}}~\mathrm{(erg)}<10^{54}$ & log-uniform\\
    $10^{-4}<n_{0}~\mathrm{(cm^{-3})}<10$ &log-uniform\\
    $2.0<p<3.5$& uniform\\
    $10^{-7}<\epsilon_B<0.50$& log-uniform\\
    $10^{-4}<\epsilon_e<0.50$& log-uniform\\
    \hline
    \end{tabular}
    \label{tab:prior}
\end{table}

\begin{table*}
    \setlength{\tabcolsep}{5pt}
    \def\arraystretch{1.25}
    \caption{Fit results for GRB 181123B for when the ATCA force-fitted data are excluded, included or the ATCA force-fitted mean is lowered by an order of magnitude. The errors on the parameters represent the 95\% credible interval.
    It can be seen that when the ATCA data are included, $\epsilon_e$ is better constrained. 
    }
    \begin{tabular}{l|r|r|r}
    \hline
    Parameter name &  \multicolumn{1}{c}{ATCA data excluded} & \multicolumn{1}{c}{ATCA data included} &  \multicolumn{1}{c}{Lower ATCA forced fit} \\
    \hline
    {\boldmath$\log_{10}E_{K, \mathrm{iso}}$} & $52.4^{+1.4}_{-1.6}        $ & $52.0^{+1.5}_{-1.2}        $ & $51.5^{+1.1}_{-0.85}       $\\
    
    {\boldmath$\log_{10}n_{0}$} & $-0.4^{+1.4}_{-1.5}        $ & $-0.5^{+1.4}_{-1.4}        $ & $-1.1^{+1.4}_{-1.3}        $\\
    
    {\boldmath$p$} & $2.92^{+0.42}_{-0.37}      $ & $2.90^{+0.42}_{-0.38}      $ & $2.97^{+0.38}_{-0.39}      $\\
    
    {\boldmath$\log_{10}\epsilon_B$} & $-2.9^{+2.5}_{-3.2}        $ & $-3.0^{+2.7}_{-3.5}        $ & $-2.5^{+2.1}_{-2.3}        $\\
    
    {\boldmath$\log_{10}\epsilon_e$} & $-1.13^{+0.82}_{-1.2}      $ & $-0.75^{+0.39}_{-0.40}     $ & $-0.60^{+0.30}_{-0.35}     $\\
    \hline
    \end{tabular}
    \label{tab:post_param}
\end{table*}

\subsubsection{Robustness of the results for more complicated models} \label{sec:robust}

The aim of the modelling analysis presented in Section~\ref{sec:mod} is to demonstrate how early-time radio data (even a non-detection) can help to constrain physical parameters in the framework of the fireball model.
However, the presence of a plateau feature in many GRB X-ray light curves indicates that energy injection and more complex emission mechanisms are at play beyond a simple forward and reverse shock. 
One of the main interpretations of X-ray plateaus observed from SGRBs 
is likely an energy injection signature from a potentially short-lived, supramassive, highly magnetised, rapidly rotating neutron star remnant, often referred to as a `magnetar' \citep[e.g.][]{zhang01,yu07,rowlinson13}. In fact, \citet{gompertz15} has performed broad-band modelling of SGRBs that includes energy injection from the spin-down of such a magnetar. Alternatively, \cite{leventis14} were able to demonstrate that X-ray plateaus could be explained by the combined emission from the reverse and forward shock, provided that the blast wave is in the thick shell regime, and such a reverse shock would also lead to an additional emission component in radio and optical wavelengths.
In the case of GRB 181123B, the X-ray light curve of the afterglow shows evidence of a plateau phase at early times that \citet{rowlinson20pp} have interpreted as energy injection from an unstable magnetar that collapsed a few hundred seconds following its formation. 
While many models have been proposed in the literature to describe the X-ray light curve behaviour of GRBs, our limited radio data-set means that an exhaustive analysis of these complex models is beyond the scope of this paper. 
Nonetheless, it is worth exploring whether the constraints we have derived from our simple forward shock model are still meaningful if we introduce additional free parameters. 

In the following we investigate how the inclusion of the ATCA data affects the posterior of modelling efforts that also include energy injection. We therefore 
incorporate energy injection into \texttt{boxfit}, 
which is modelled by varying the isotropic equivalent kinetic energy ($E_{K, \mathrm{iso}}$) as a power-law in time. In this case, $E_{K, \mathrm{iso}}$ is described as:

\begin{equation}
    E = 
    \begin{cases} 
      E_0 & t \leq t_{\mathrm{inj}} \\
      E_0 (t / t_{\mathrm{inj}})^\alpha &  t_{\mathrm{inj}} < t \leq t_{\mathrm{inj}} + dt_{\mathrm{inj}} \\
      E_0 (1 + dt_{\mathrm{inj}} / t_{\mathrm{inj}})^\alpha & t_{\mathrm{inj}} + dt_{\mathrm{inj}} < t 
   \end{cases}
\end{equation}

\noindent where the three additional parameters are defined as:

\begin{itemize}
  \item $t_{\mathrm{inj}}$: Start time of the energy injection in seconds (s).
  \item $dt_{\mathrm{inj}}$: Duration of energy injection in seconds (s).
  \item $\alpha$: Power-law index of the energy injection.
\end{itemize}

\noindent The assumed prior distributions for the energy injection parameters can be seen in Table~\ref{tab:prior_ei}. We use the same priors as before for all other burst parameters (Table~\ref{tab:prior}).

\begin{figure}
  \includegraphics[width=.45\textwidth]{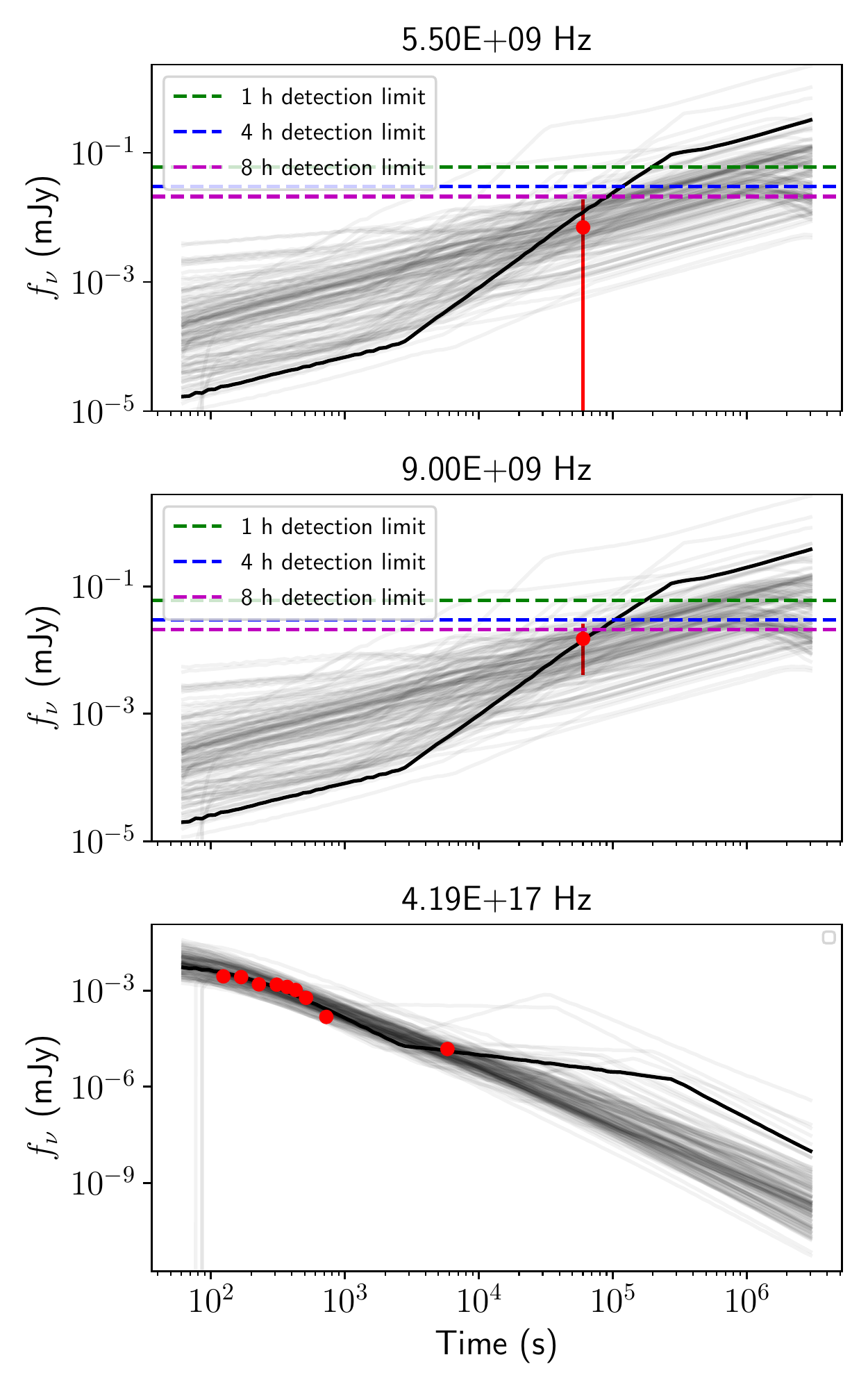}
  \caption{As for Figure~\ref{fig:lc_post} but with energy injection incorporated into our modelling. 
  } 
  \label{fig:lc_post_ei}
\end{figure}

The resulting best-fit parameters for the afterglow modelling of GRB 181123B that includes energy injection for the cases when the ATCA force-fitted flux densities are included and excluded can be found in Table~\ref{tab:post_param_ei}, with the resulting light curves for the posterior predictive distribution, together with the best fit model plotted in Figure \ref{fig:lc_post_ei}. 
Due to the limited multi-wavelength coverage of the afterglow, we are not able to place tight constraints on the energy injection parameters. 
However, note that $\epsilon_e$ continues to be well constrained when ATCA data are included, even with a more complex model with additional parameters (see Figure~\ref{fig:marge_comp_ei}, which shows the marginal distributions of the parameters for the energy injection cases when ATCA data are included and excluded). 
In addition, Figure~\ref{fig:marge_comp_classic_ei} compares the obtained marginalized distribution for the GRB parameters (omitting the energy injection parameters) for the two modelling cases that include the ATCA data: with and without including energy injection. It can be seen that the resulting parameter values common between both fits are consistent within the 95\% credible intervals despite different model complexities.
However, while the model is consistent with the \swift{}-UVOT upper-limits, it over-predicts the $i$-band detection \citep{paterson20}, which demonstrates the limitations of our modelling.

The modelling of GRB 181123B shows that early-time radio observations, 
regardless of whether they are detections or non-detections,
are able to constrain the fraction of thermal energy in the accelerated electrons, $\epsilon_e$, beyond what is possible with just the \swift-XRT X-ray light curve data. 
Our modelling also predicted that observations at later times (1-10 days) may have resulted in a detection of the forward shock, which would further constrain the GRB parameters.

\begin{table}
    \setlength{\tabcolsep}{5pt}
    \def\arraystretch{1.25}
    \caption{Assumed priors for the energy injection parameters for all modelling efforts.}
    \begin{tabular}{l|r}
    \hline
    \multicolumn{1}{l}{Parameter range} & \multicolumn{1}{r}{Prior distribution} \\
    \hline
    $8.64<t_{\mathrm{inj}}~\mathrm{(s)}<4.32 \times 10^5$ & log-uniform\\
    
    $8.64<dt_{\mathrm{inj}}~\mathrm{(s)}<4.32 \times 10^5$ & log-uniform\\
    
    $0.0<\alpha<2.0$& uniform \\
    \hline
    \end{tabular}
    \label{tab:prior_ei}
\end{table}

\begin{table}
    \setlength{\tabcolsep}{5pt}
    \def\arraystretch{1.25}
    \caption{Fit results for GRB 181123B for when the ATCA force-fitted data are excluded and included, with 
    energy injection also included in our modelling. The errors on the parameters represent the 95\% credible interval. It can be seen that when the ATCA data are included, $\epsilon_e$ is better constrained. 
    }
    \begin{tabular}{l|r|r}
    \hline
    Parameter name &  \multicolumn{1}{c}{ATCA data excluded} & \multicolumn{1}{c}{ATCA data included} \\
    \hline
    {\boldmath$\log_{10}E_{K, \mathrm{iso}}$} & $51.8^{+1.9}_{-1.3}        $   & $51.4^{+1.7}_{-0.91}       $\\
    
    {\boldmath$\log_{10}n_{0}$}    & $-0.7^{+1.6}_{-1.7}        $ & $-0.9^{+1.7}_{-1.8}        $ \\
    
    {\boldmath$p              $} &$3.08^{+0.42}_{-0.47}      $ & $3.16^{+0.35}_{-0.50}      $\\
    
    {\boldmath$\log_{10}\epsilon_{B}$} &$-1.9^{+1.6}_{-3.5}        $ & $-2.1^{+1.9}_{-4.0}        $\\
    
    {\boldmath$\log_{10}\epsilon_{e}$} &$-1.01^{+0.70}_{-1.2}      $ & $-0.66^{+0.35}_{-0.39}     $\\
    
    {\boldmath$\log_{10}t_{\mathrm{inj}}$ (days)} &$-1.23^{+1.7}_{-0.91}      $ & $-1.32^{+1.8}_{-0.84}      $\\
    
    {\boldmath$\log_{10}dt_{\mathrm{inj}}$ (days)} & $-1.2^{+1.8}_{-2.3}        $ & $-1.1^{+1.8}_{-2.3}        $\\
    
    {\boldmath$\alpha         $} &
    $0.86^{+0.94}_{-0.77}      $ & $0.88^{+0.92}_{-0.76}      $\\
    \hline
    \end{tabular}
    \label{tab:post_param_ei}
\end{table}

\begin{figure}
  \includegraphics[width=.45\textwidth]{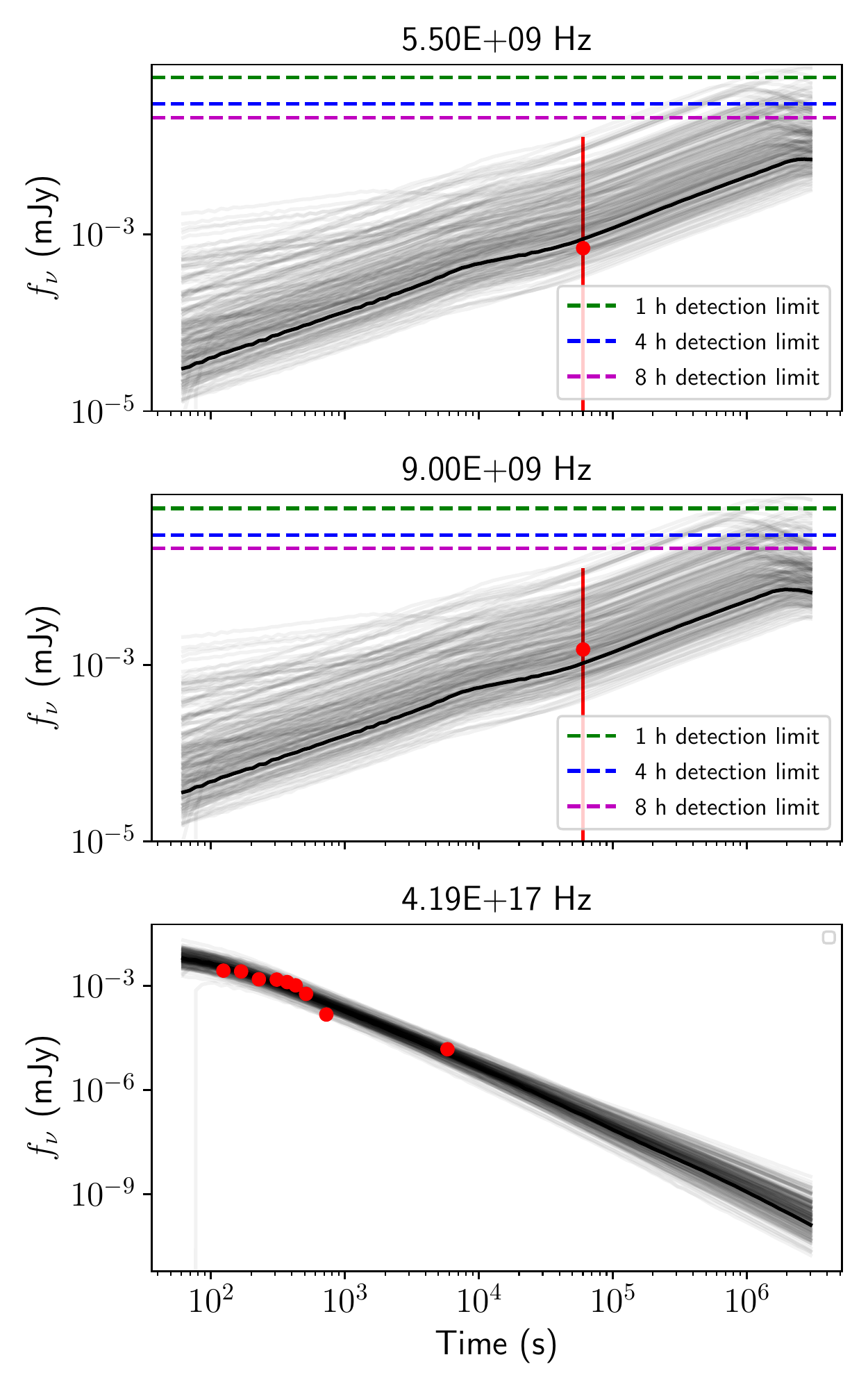}
  \caption{As for Figure~\ref{fig:lc_post} but with the ATCA forced-fitted flux densities lowered by an order of magnitude. 
  }
  \label{fig:lc_post_mean}
\end{figure}

To test how the dependencies between the parameters in our model are affected by 
the mean of the forced fitted values, we ran an additional fit using \texttt{boxfit} (without energy injection) where we lowered the mean of the ATCA force-fitted flux density values by an order of magnitude. Table~\ref{tab:post_param} shows the resulting inferred parameters with the best fit light curves shown in Figure~\ref{fig:lc_post_mean}. 
The main effect of lowering the mean value of the force-fitted ATCA flux densities on the parameters was to decrease the mean value of the circumburst medium density ($n_0$) and increase the fraction of thermal energy in the  magnetic fields ($\epsilon_B$) by a similar amount (factor of $\sim3-4$) when compared to the results from fitting the original force-fitted ATCA flux densities.
However, note that all parameters presented in Table~\ref{tab:post_param} that were derived from including ATCA force-fitted flux densities in our  modelling still agree within the 95\% credible intervals of the modelling performed without the ATCA information.
The reduction in the ATCA force-fitted mean flux densities also indicate that the predicted peak forward shock emission in the radio band would be delayed and also drop below the ATCA $3\sigma$ detection limits (see Figure~\ref{fig:lc_post_mean}) when compared to the fit performed with the measured force-fitted flux densities shown in Figure~\ref{fig:lc_post}.

Overall, our analysis demonstrates the importance of quoting force-fitted flux density values for radio transients over just reporting $3\sigma$ upper-limits, which is traditional in most fields of astrophysics. If only upper-limits are reported then it is not always possible to incorporate this information into some modelling analyses (e.g. like our GRB afterglow modelling), which means we are throwing away important data that could further constrain the physics of an event or source. 
Finally, the inclusion of early-time radio force-fitted flux densities allows us to make predictions about the time and brightness of the forward shock peak in the radio band, which can inform late-time radio follow-up strategies. 

\subsection{Comparisons of GRB 181123B to radio-detected SGRBs}\label{sec:comp}

We now compare our parameter constraints on the micro- and macro-physical properties of GRB 181123B resulting from our modelling to those obtained for other radio-detected GRBs. For this comparison, we only focus on the parameters derived from our forward-shock modelling using \texttt{boxfit} with the inclusion of the ATCA data (see Table~\ref{tab:post_param}). 
Six of the eight radio-detected SGRBs have constraints on the same parameters (e.g. Table~\ref{tab:prior}) through afterglow modelling 
\citep[GRB 050724A, 051221A, 130603B, 140903A, 160821B, GRB 200522A;][]{berger05,soderberg06,fong14,pandey19,troja16,zhang17,troja19,lamb19,fong21}. 
The advantage of these modelling efforts was that these events had extensive multi-wavelength data (radio, infrared, optical, X-ray) so the afterglow analysis 
led to stringent constraints on these parameters and in many cases, an estimate of 
the gamma-ray jet opening angle, which has important implications for rate calculations. 
While the parameters we derived for GRB 181123B were far less constrained, our 95\% credible intervals agree with those values derived for the six  SGRBs mentioned above, but the upper limits in our prediction for both $n_0$ and $p$ are much higher overall. 
In fact, our derived accelerated electron distribution is steeper than that usually expected for GRBs \textbf{($p=2.90^{+0.42}_{-0.38}$)} but not unreasonably so.
Overall, as previously mentioned, the inclusion of ATCA force-fitted flux densities within 1 day post-burst have allowed us to place 
reasonable constraints on $\epsilon_{e}$. 

\section{Conclusions}

In this paper, we introduce the ATCA rapid-response observing mode by presenting results from the first successful SGRB trigger on GRB 181123B. This new mode of operations allows the telescope to automatically and rapidly respond to transient alerts broadcast via VOEvents, causing the active observing programme to be interrupted to allow for time-critical observations of transient phenomena. Successful triggers on LGRBs (see Section~\ref{sec:trig_per}) have demonstrated that if the source is above the horizon, the ATCA can be on target and observing the event as fast as 3 minutes post-burst, allowing us to probe this early-time radio regime over a wide range of frequencies ($1.1-25$\,GHz) with full polarisation, and in a variety of array configurations.

The ATCA rapid-response observations of GRB 181123B began 12.6\,hr post-burst, as soon as the target had risen above the horizon, collecting 8.3\,h of data at 5.5 and 9\,GHz. While no radio emission was detected from GRB 181123B, we quote force-fitted flux densities, which enabled more constraining GRB afterglow modelling to be performed then would usually be possible with just X-ray data from \swift{}-XRT. The addition of early-time radio data in the modelling allowed us to obtain more stringent constraints on the fraction of thermal energy in the electrons behind the shock wave ($\epsilon_e$), which in turn allowed us to predict the peak in the forward shock radio afterglow emission around $\sim10$\,days post-burst.  
This modelling indicates that $\gtrsim3-4$\,days post-burst, the radio afterglow of GRB 181123B may have been detectable with a $\geq4$\,hr ATCA integration. 

Our results demonstrate the importance of including early-time radio observations in afterglow modelling efforts for constraining GRB blast wave properties and predicting the late-time peak in the radio forward shock, regardless of whether or not there is a detection, provided that force-fitted flux densities are quoted rather than upper-limits in the case of a non-detection. 

This project also demonstrates the importance of implementing rapid-response observing systems on radio telescopes to probe a new parameter space in transient science. 
Early-time radio observations of SGRBs can allow us to distinguish between different sources of early-time synchrotron afterglow emission \citep[e.g.][]{kyutoku14} and even detect prompt and persistent coherent signals predicted to be produced during the compact merger and from the merger remnant \citep[e.g.][]{rowlinsonanderson19}.
Detections from rapid-response radio observations will provide crucial insight into the radio brightness and timescales we might expect from aLIGO/Virgo-detected merging BNSs or NS-BH systems, which will aid in our search for electromagnetic counterparts in the large GW positional error regions.
Other science cases include 
LGRBs, which may show bright, early-time radio emission from the reverse shock \citep[e.g.][]{anderson14} and flare stars, which have shown simultaneous high-energy and radio flaring behaviour \citep[e.g.][]{fender15}.

Finally, our efforts running rapid-response programs on ATCA act as an excellent test for transient observing strategies with the SKA. 
It is only through the utilisation of a rapid-response system that we can exploit the SKA to study early-time BNS and BH-NS merger physics down to sub-micro-Jansky levels. 

\section*{acknowledgements}
We thank the referee for their careful reading of the manuscript and recommendations.
GEA is the recipient of an Australian Research Council Discovery Early Career Researcher Award (project number DE180100346) and JCAM-J is the recipient of Australian Research Council Future Fellowship (project number FT140101082) funded by the Australian Government. 

The Australia Telescope Compact Array (ATCA) is part of the Australia Telescope National Facility, which is funded by the Australian Government for operation as a National Facility managed by CSIRO. 
This work made use of data supplied by the UK {\it Swift} Science Data Centre at the University of Leicester and the {\it Swift} satellite. {\it Swift}, launched in November 2004, is a NASA mission in partnership with the Italian Space Agency and the UK Space Agency. {\it Swift} is managed by NASA Goddard. Penn State University controls science and flight operations from the Mission Operations Center in University Park, Pennsylvania. Los Alamos National Laboratory provides gamma-ray imaging analysis.

The ATCA rapid-response front-end software makes use of {\sc comet} \citep{swinbank14} and
{\sc voevent-parse} \citep{staley16pp} to process the incoming VOEvents.
Both the front-end ({\sc vo\_atca})\footnote{https://github.com/mebell/vo\_atca} and back-end ({\sc atca-rapid-resonse-api})\footnote{https://github.com/ste616/atca-rapid-response-api} software for ATCA rapid-response triggering rely on the {\sc Astropy}, a community-developed core Python package for Astronomy \citep{TheAstropyCollaboration2013,TheAstropyCollaboration2018}, {\sc numpy} \citep{vanderWalt_numpy_2011} and {\sc scipy} \citep{Jones_scipy_2001} python modules. This research also makes use of {\sc matplotlib} \citep{hunter07}. 
This research has made use of NASA's Astrophysics Data System. 
This research has made use of SAOImage DS9, developed by Smithsonian Astrophysical Observatory.
This research has made use of the VizieR catalogue access tool \citep{ochsenbein00} and the SIMBAD database \citep{wenger00}, operated at CDS, Strasbourg, France.

\section*{Data Availability Statement}
The unprocessed ATCA visibility datasets are public and can be accessed via the ATCA online archive (\href{https://atoa.atnf.csiro.au/}{https://atoa.atnf.csiro.au/}) under project code C3204 and PI G Anderson.
The reduced images of these observations have been uploaded to Zenodo \citep{anderson21zen}.

\bibliographystyle{mnras}
\bibliography{papers2}

\appendix

\section{Marginalised parameter distributions for different model fits}\label{sec:app_A}

The following Figures~\ref{fig:marge_comp}, \ref{fig:marge_comp_ei} and \ref{fig:marge_comp_classic_ei} show the marginalised parameter distributions for different modelling tests performed in Section~\ref{sec:mod}.

\begin{figure*}
  \begin{center}
  \includegraphics[width=\textwidth]{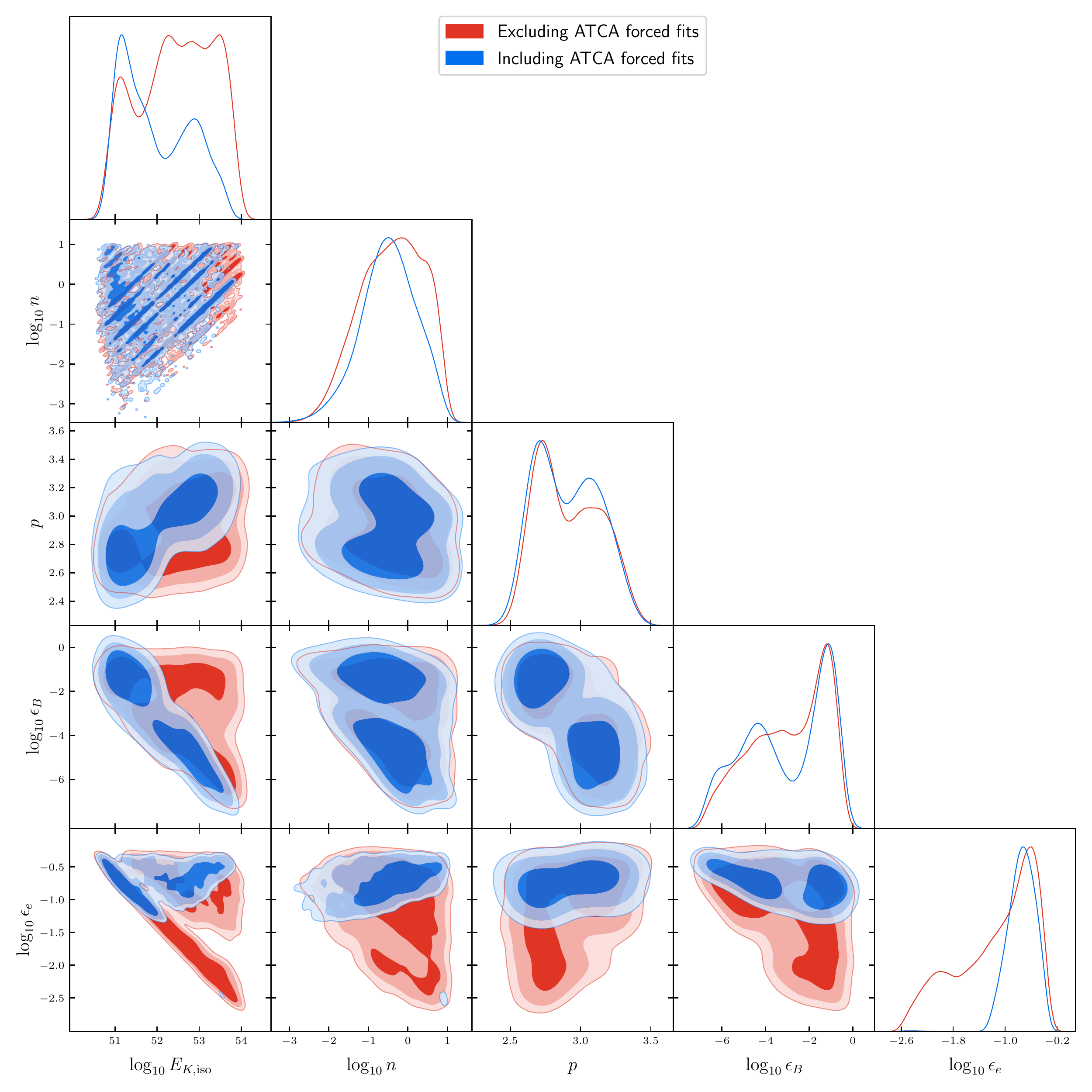}
  \caption{The corner plot for the marginalized parameter distributions resulting from forward-shock modelling of the GRB 181123B afterglow using \texttt{boxfit}. The red distributions show the modelling results when the ATCA force-fitted flux densities are not included in the data set and the blue distributions show the results for 
  when the ATCA data are included.
  }
  \label{fig:marge_comp}
  \end{center}
\end{figure*}

\begin{figure*}
  \begin{center}
  \includegraphics[width=\textwidth]{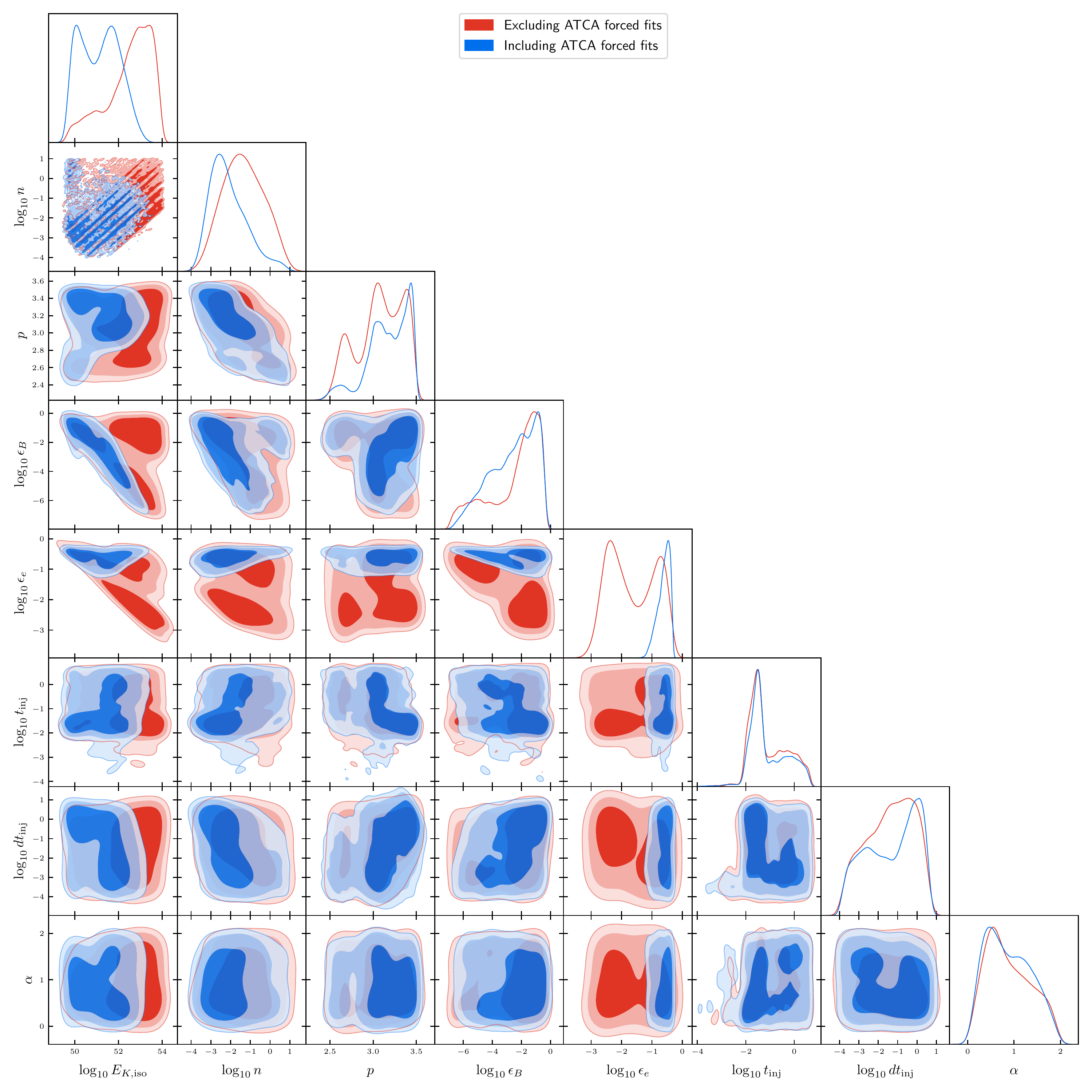}
  \caption{As for Figure~\ref{fig:marge_comp} but with energy injection included in the model. 
  }
  \label{fig:marge_comp_ei}
  \end{center}
\end{figure*}

\begin{figure*}
  \begin{center}
  \includegraphics[width=\textwidth]{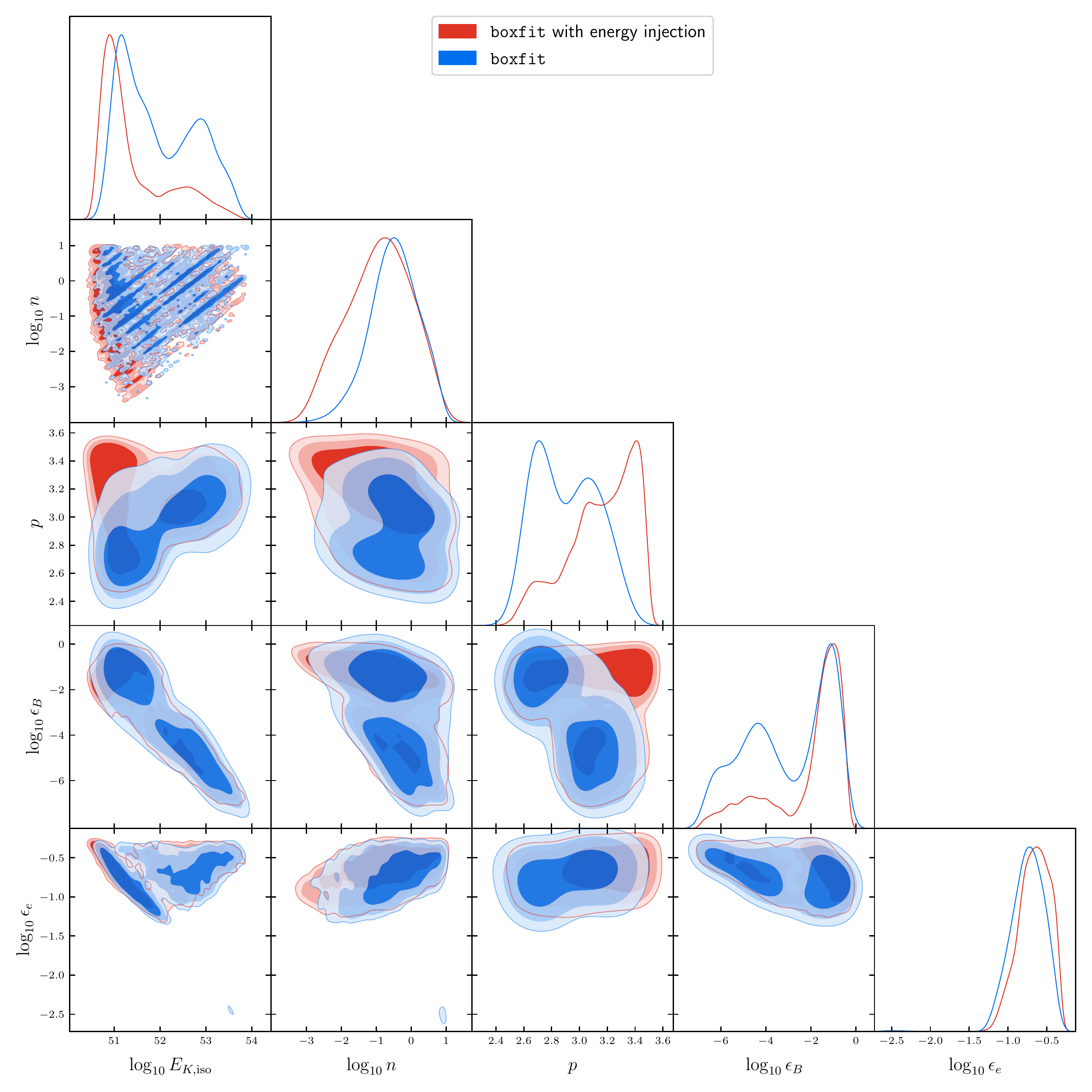}
  \caption{As for Figure~\ref{fig:marge_comp}, however, in this case both presented parameter distributions include the ATCA data. 
  The red distributions show the modelling results when energy injection is included in the model and the blue distributions show the results without the inclusion of energy injection. Although including energy injection increases the complexity of the model, it can be seen that $\epsilon_e$ is well constrained in either case when ATCA data are included. The additional parameters for the energy injection case ($t_{\mathrm{inj}}$, $dt_{\mathrm{inj}}$ and $\alpha$) are omitted in this plot for clarity.
  }
  \label{fig:marge_comp_classic_ei}
  \end{center}
\end{figure*}

% Don't change these lines
\bsp	% typesetting comment
\label{lastpage}
\end{document}